\documentclass[onecolumn,preprintnumbers,elsart]{revtex4}
\usepackage{amssymb}
\usepackage{amsmath}
\usepackage{graphicx}
\usepackage{dcolumn}
\usepackage[center]{subfigure}
\usepackage{array}
\usepackage{booktabs}
\usepackage{multirow}
\usepackage{color}

\begin{document}

\title{Two-dimensional vortex quantum droplets get thick}
\author{Zeda Lin$^{1,\S }$, Xiaoxi Xu $^{1,\S }$, Zikang Chen$^{1}$, Ziteng
Yan$^{1}$, Zhijie Mai$^{2,3}$, Bin Liu$^{1}$}
\email{binliu@fosu.edu.cn}
\affiliation{$^{1}$School of Physics and Optoelectronic Engineering, Foshan University,
Foshan 528000, China}
\affiliation{$^{2}$ Department of Applied Physics, College of Electronic Engineering,
South China Agricultural University, Guangzhou 510642, China}
\affiliation{$^{3}$ School of Mathematics, Statistics and Physics, Newcastle University,
Newcastle upon Tyne, United Kingdom, NE1 7RU}
\affiliation{$^{\S }$ These authors contributed equally to this work.}

\begin{abstract}
We study two-dimensional (2D) vortex quantum droplets (QDs) trapped by a
thicker transverse confinement with $a_{\perp }>1\mu m$. Under this
circumstance, the Lee-Huang-Yang (LHY) term should be described by its
original form in the three-dimensional (3D) configuration. Previous studies
have demonstrated that stable 2D vortex QDs can be supported by a thin
transverse confinement with $a_{\perp }\ll 1\mu m$. In this case, the LHY
term is described by a logarithm. Hence, two kinds of confinement features
result in different mechanisms of the vortex QDs. The stabilities and
characteristics of the vortex QDs must be re-identified. In the current
system, we find that stable 2D vortex QDs can be supported with topological
charge number up to at least $4$. We reformulated their density profile,
chemical potential and threshold norm for supporting the stable vortex QDs
according to the new condition. Unlike the QDs under thin confinement, the
QDs in the current system strongly repel each other because the LHY term
features a higher-order repulsion than that of the thin confinement system.
Moreover, elastic and inelastic collisions between two moving vortex QDs are
studied throughout the paper. Two kinds of collisions can be characterized
by exerting different values of related speed. The dynamics of the stable
nested vortex QD, which is constructed by embedding one vortex QD with a
smaller topological number into another vortex QD with a larger number of
topological charge, can be supported by the system. \newline
\end{abstract}

\maketitle

\section{Introduction}

Recently, a novel type of quantum liquid self-bound state, referred to as
quantum droplets (QDs), has been formed with the help of the zero-point
quantum fluctuations of the collective Bogoliugov mode, which can be
described theoretically by the Lee-Huang-Yang (LHY) correction \cite{LHY1957}
in three-dimensional (3D) space. This correction describes a repulsive
beyond mean-field (MF) force, which plays the role of a higher-order
nonlinear repulsive term and can arrest the collapse of attractive Bose
gases induced by the MF force. These attractive Bose gases are collapsing
dipolar Bose-Einstein condensates with a strong attractive dipole-dipole
interaction \cite{Saito2016,PRL116_215301,NJP21_093027}, and a collapsing
Bose-Bose (BB) mixture when the interspecies attraction is stronger than the
intraspecies repulsion \cite{Petrov2015}. Experiments have reported that
self-bound QDs were formed by the dipolar Bose-Einstein condensates (BECs)
of dysprosium, erbium \cite{Schmitt2016,Chomaz2016,PRL120_160402}, and the
BB mixture of kalium \cite{Cheiney2018,Cabrera2017,PRL120_235301}.

In the case of the BB mixture, the LHY correction can manifest a different
mechanism for the lower dimensional system \cite%
{Petrov2016,PRA99_051601,PRA98_051603,PRA98_051604,PRA97_063605}. In the
two-dimensional (2D) system with thin confinement with $a_{\perp }\ll 1\mu m$%
, the energy of the LHY term is $\sim n^{2}\ln (n/\sqrt{e})$ (where $n$
dominates the density of the QD) \cite{cluster,PRL113_160405,YLi2017} and
gives rise to a term including a logarithm in the Gross-Pitaevskii equation
(GPE). According to the behavior of the logarithm, the LHY term manifests
repulsion and attraction when the QDs are under extremely dense and dilute
conditions, respectively. The reduction of the BB mixture with the LHY
corrections to the one-dimensional (1D) system is drastically different.
Under this circumstance, the QDs are formed when the LHY term features only
attraction, contrary to its repulsive sign in higher dimensions, while the
total effect of the cubic mean-field force is tuned to repulsion, which
competes with the LHY-induced attraction \cite%
{Astrakharchik2018,Bin2019,Zhouzheng}. Recent studies have revealed that the
LHY term in the lower dimensional systems can stabilize the QD embedded with
vortices. It has been reported that vortex QDs can be stabilized up to at
least $S=5$ in a 2D thin confinement system \cite{Yongyao2018}, and
semidiscrete vortex QD, which can also be stabilized up to at least $S=5$,
can be created in arrays of the coupled 1D confinement system \cite{Xiliang}%
. In contrast with these results, vortex QDs in single-component dipolar
condensates were found to be unstable \cite{Macri}. However, for the case of
2D vortex QDs in the BB mixture, the stable 2D vortex QDs are created based
on the condition of thin confinement in the transverse direction (i.e., $%
a_{\perp }\ll 1\mu m$). As mentioned above, the LHY term contains a
logarithm under this circumstance. If the transverse confinement thickens
(i.e., $a_{\perp }>1\mu m$), which is also the natural scale for BEC
confinement in the 2D configuration, the LHY term may retreat to its
original form, which is the same as the form in the 3D configuration (i.e.,
a quartic term \cite{Petrov2015,PRL112_090401}). In this case, the LHY term
manifests only strong repulsion, which is different from the effect of the
logarithm in the case of thin transverse confinement. It is necessary to
point out that stabilization of a vortex QD for the BB mixture in the full
3D configuration remains a challenging issue \cite{YVK2018}. Because the LHY
term plays an important role in stabilizing the vortex QD, how the change of
the LHY term affects the stabilities and characteristics of the vortex QD is
worth exploring.

In this paper, we aim to reconsider the stabilities and characteristics of
2D vortex QDs under thicker transverse confinement with the value of $%
a_{\perp }$ within a few microns. Under this circumstance, the system can
still be termed as a quasi-2D one. The LHY term should be replaced by a
quartic term in the GPE, which is the same as its original form in the 3D
configuration. The rest of this paper is structured as follows: the model
for the current system is described in Section II, and results of the 2D
vortex QD in the new condition are discussed in Section III, and this work
is concluded in Section IV.

\section{The model}

According to the ref. (\cite{CM_boris,arxiv_czp}), the underlying 3D GP
equation supplemented by the LHY-induced quartic self-repulsion term can be
written, in the scaled form, as%
\begin{equation}
i{\frac{\partial }{\partial t}}\Phi =-{\frac{1}{2}}\nabla _{3D}^{2}\Phi
-g|\Phi |^{2}\Phi +|\Phi |^{3}\Phi +V\left( z\right) \Phi ,  \label{3D_GPE}
\end{equation}%
where $\Phi $\ stands for equal wave functions of two components of the
BECs, and $V\left( z\right) =z^{2}/(2a_{\perp }^{4})$\ is the transverse
confinement. Then, the 3D to 2D reduction is performed by means of the
standard substitution, as follows,%
\begin{equation}
\Phi \left( x,y,z,t\right) =\Psi \left( x,y,t\right) \exp \left( -\frac{1}{%
2a_{\perp }^{2}}z^{2}-i\frac{t}{2a_{\perp }^{2}}\right) .  \label{trans1}
\end{equation}%
\ Substituting the solution Eq. (\ref{trans1}) into Eq. (\ref{3D_GPE}), and
followed by the averaging of Eq. (\ref{3D_GPE}) in the transverse direction,
one yields%
\begin{equation*}
i{\frac{\partial \Psi }{\partial t}}=-{\frac{1}{2}}\nabla _{2D}^{2}\Psi -%
\frac{g}{\sqrt{3}}|\Psi |^{2}\Psi +\frac{1}{2}|\Psi |^{3}\Psi .
\end{equation*}%
Further, with the help of additional rescaling, $(x,y)\rightarrow \sqrt{2}%
(x,y)$, $t\rightarrow 2t$, $g\rightarrow \frac{\sqrt{3}}{2}g$, the effective
2D GPE can be written as follows:%
\begin{equation}
i\frac{\partial \Psi }{\partial t}=-{\frac{1}{2}}\nabla _{2D}^{2}\Psi
-g|\Psi |^{2}\Psi +|\Psi |^{3}\Psi .  \label{GPE}
\end{equation}%
Where $g>0$ is the strength of the self-attractive cubic nonlinearity (the
intercomponent attraction being slightly stronger than the repulsion for
each component \cite{Petrov2015}). The total norm under the symmetry
condition can be characterized as
\begin{equation}
N=\int \int |\Psi (\mathbf{r})|^{2}dxdy.  \label{Norm}
\end{equation}%
and the Hamiltonian (energy) corresponding to Eq. (\ref{GPE}) is
\begin{equation}
E={\frac{1}{2}}\int \int \left[ |\nabla \Psi |^{2}-g|\Psi |^{4}+{\frac{4}{5}}%
|\Psi |^{5}\right] dxdy.  \label{Ham}
\end{equation}

The objective of this work is to make use of the LHY effect for the
stabilization of the vortex QDs in the present system. Thus the stationary
QDs solutions with topological charge number $S=1$, $2$, $\cdots $ in the
polar coordinates are looked for
\begin{equation}
\Psi (\mathbf{r},t)=\phi (r)\exp (-i\mu t+iS\theta ),  \label{QD}
\end{equation}%
Substituting the solution Eq. (\ref{QD}) into Eq. (\ref{GPE}), the real
amplitude function $\phi (r)$ obeys a radial equation%
\begin{equation}
\mu \phi =-{\frac{1}{2}}\left( {\frac{d^{2}}{dr^{2}}}+{\frac{1}{r}}{\frac{d}{%
dr}}-{\frac{S^{2}}{r}}\right) \phi -g\phi ^{3}+\phi ^{4}.  \label{phir}
\end{equation}

The stability of the stationary solution is analyzed by means of the
linearized Bogoliugov--de Gennes (BdG) equations for perturbed wave
functions, taken as
\begin{equation}
\Psi (\mathbf{r},t)=\left[ \phi (r)+w(r)e^{-i\lambda t+im\theta }+v^{\ast
}(r)e^{i\lambda ^{\ast }t-im\theta }\right] e^{-i\mu t+iS\theta },
\label{PerSolution}
\end{equation}%
where $w$, $v$, and $\lambda $ are eigenmodes and the instability growth
rate corresponding to an integer azimuthal index $m$ of the perturbation.
The linearization around the stationary solution leads to equations
\begin{eqnarray}
&&\lambda w=-{\frac{1}{2}}\left[ {\frac{d^{2}}{dr^{2}}}+{\frac{1}{r}}{\frac{d%
}{dr}}-{\frac{(S+m)^{2}}{r^{2}}}\right] w+\left( -2g\phi ^{2}+{\frac{5}{2}}%
\phi ^{3}-\mu \right) w+\left( -g\phi ^{2}+{\frac{3}{2}}\phi ^{3}\right) v,
\notag \\
&&\lambda v={\frac{1}{2}}\left[ {\frac{d^{2}}{dr^{2}}}+{\frac{1}{r}}{\frac{d%
}{dr}}-{\frac{(S-m)^{2}}{r^{2}}}\right] v-\left( -2g\phi ^{2}+{\frac{5}{2}}%
\phi ^{3}-\mu \right) v-\left( -g\phi ^{2}+{\frac{3}{2}}\phi ^{3}\right) w.
\end{eqnarray}%
Numerical solution of the linearized equations produces a spectrum of
eigenfrequencies $\lambda $, the stability condition being that the spectrum
of $\lambda $ must be real for at least $m=0,1,2,3$ \cite{Mihalache,Nir}.
Moreover, the stability of the stationary solutions are also verified by
direct simulations of the perturbed evolution in the framework of Eq. (\ref%
{GPE}).

\section{Results and discussion}

\subsection{Stationary solutions}

\begin{figure}[h]
{\includegraphics[width=1.0\columnwidth]{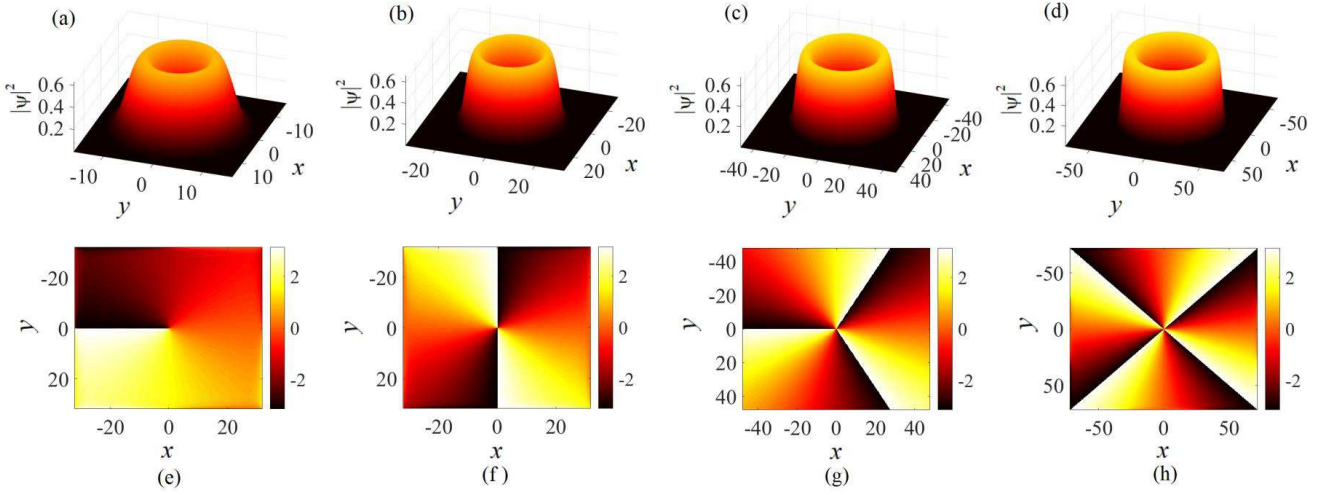}}
\caption{Typical examples of stable vortex QDs with $S=1$ to $4$ (from left
to right). (a-d) Density patterns of the vortex QDs with $(N,S)=(200,1)$, $%
(N,S)=(500,2)$, $(N,S)=(1200,3)$, and $(N,S)=(2500,4)$. (e-h) The
corresponding phase diagrams of the vortex QDs are in panels (a-d),
respectively. }
\label{Exp}
\end{figure}

Stationary solutions for Eq. (\ref{GPE}) are numerically solved by the
imaginary time method \cite{Chiofalo,Jianke}. Stable vortex QDs in this 2D
system with thicker transverse confinement are found when the topological
charge $S=1$, $2$, $3$, and $4$. Typical examples of the density pattern as
well as the phase diagram of these vortex QDs for $g=1$ are displayed in
Fig. \ref{Exp}. In Fig. 2, we give the direct simulations of the perturbed
evolution results of $S=1$\ and $S=4$, respectively, as well as the
perturbation eigenvalues for the corresponding vortex QDs with $S=1$\ and $%
S=4$\ with different azimuthal index $m$. These results demonstrate that the
vortex QDs can be stable at least up to $S=4$. Further, Figs. \ref{Aeff}%
(a-d) show the density pattern of vortex QDs of $S=1$ with different values
of $N$, which indicates that the vortex QDs in this system are also flat-top
for sufficiently large Norm value.

To study the characteristics of the vortex QDs, we define the effective area
for the QDs as
\begin{equation}
A_{\mathrm{eff}}={\frac{\left( \int \int |\Psi |^{2}dxdy\right) ^{2}}{\int
\int |\Psi |^{4}dxdy}},  \label{Aeffeq}
\end{equation}%
The functions of $A_{\mathrm{eff}}$ for the vortex QDs with $(S,g)=(1,1)$
are shown in Fig. \ref{Aeff}(e), which indicates that the curve for $A_{%
\mathrm{eff}}(N)$ expands linearly with the increase of the total norm. The
curve of $A_{\mathrm{eff}}(N)$ can be linearly fitted by $A_{\mathrm{eff}%
}=125+1.5N$. In Fig. \ref{Aeff}(f), the energy of the vortex QDs with $S=1$\
as a function of $N$\ are given, which shows that $E$ decrease linearly with
the increase of $N$.

\begin{figure}[h]
{\includegraphics[width=0.7\columnwidth]{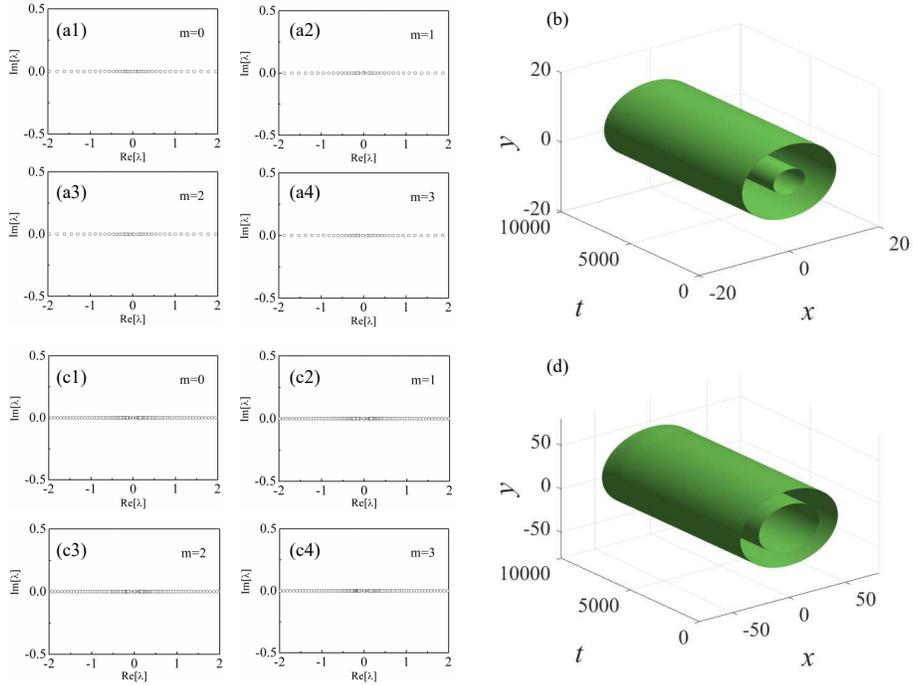}}
\caption{Perturbation eigenvalues for the corresponding vortex QDs with $S=1$
and $N=200$ for different azimuthal index $m$ [in panels (a1)-(a4)] and
direct simulations of the perturbed evolution of $\Psi $ [in panel (b)] are
displayed, respectively. Perturbation eigenvalues for the corresponding
vortex QDs with $S=4$ and $N=2500$ for different azimuthal index $m$ [in
panels (c1)-(c4)] and direct simulations of the perturbed evolution of $\Psi
$ [in panel (d)] are displayed, respectively. }
\label{Exp1}
\end{figure}
\begin{figure}[h]
{\includegraphics[width=0.8\columnwidth]{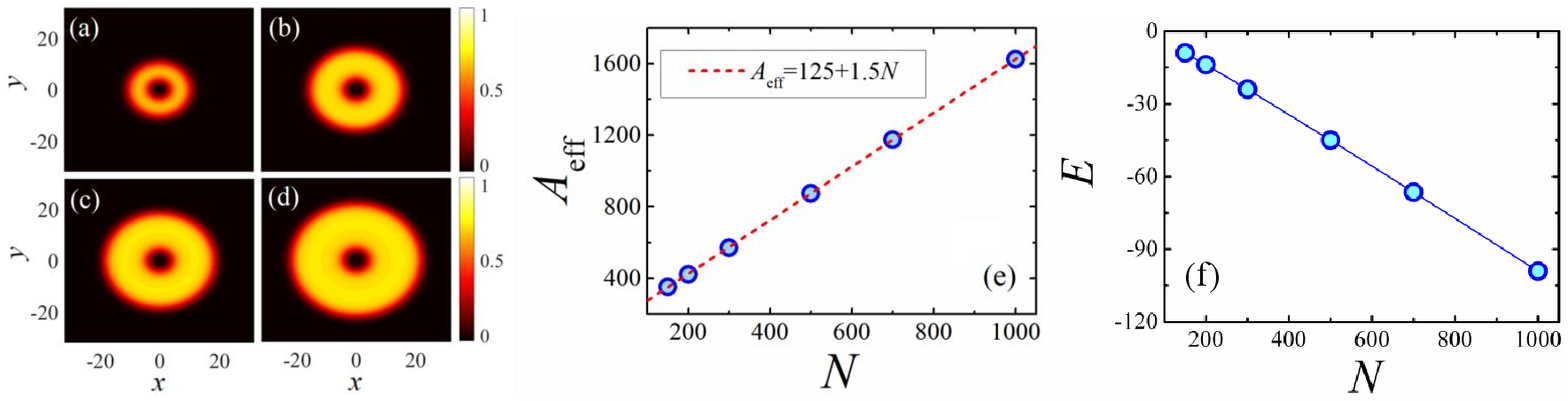}}
\caption{(a-d) Typical examples of stable vortex QDs with $(N,S)=(200,1)$, $%
(500,1)$, $(700,1)$, and $(1000,1)$, respectively. (e) The effective area of
the vortex QDs with $S=1$ as a function of $N$. (f) The energy of the vortex
QDs with $S=1$ as a function of $N$. }
\label{Aeff}
\end{figure}

\begin{figure}[h]
{\includegraphics[width=0.8\columnwidth]{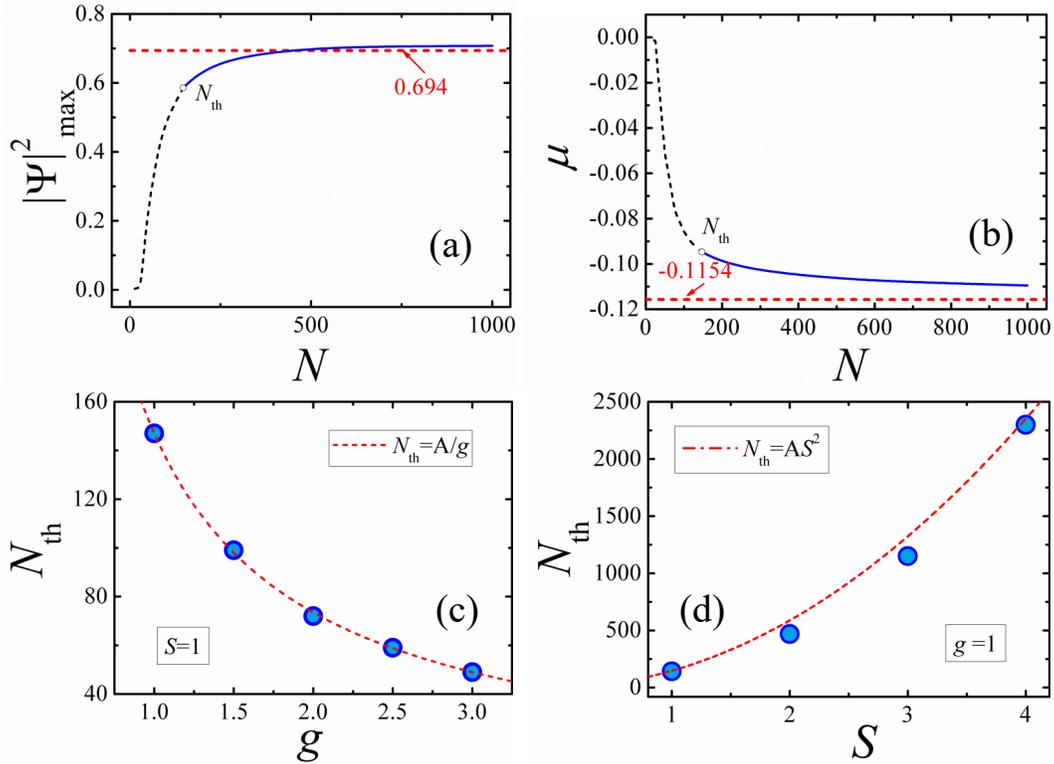}}
\caption{(a) Peak density of the vortex QD, $|\Psi |_{\max }^{2}$, versus $N$%
. (b) Chemical potential of the vortex QDs as a function of $N$. In panels
(a,b), we select $(g,S)=(1,1)$. (c) Threshold value of total norm for vortex
QD with $S=1$ as a function of $g$. (d) Threshold value of total norm for
vortex QD with $g=1$ versus topological charge number $S$.}
\label{character}
\end{figure}

The influences of the total norm on other characteristics of the vortex QDs
are also studied. Figs. \ref{character}(a,b) display the peak value, $|\phi
|_{\max }^{2}$, and the chemical potential, $\mu $, as the functions of $N$.
The function $|\phi |_{\max }^{2}(N)$ manifests that the curve of the
function saturates to a fixed value for a sufficiently large value of $N$.
Furthermore, the function of $\mu (N)$ satisfies the Vakhitov-Kolokolov (VK)
criterion, i.e., $d\mu /dN<0$, which is a necessary stability condition for
self-trapping modes in the attractive media \cite{VK}.

Similar to the case of vortex QDs in the thin confinement system, the
numerical simulations also find a threshold norm, $N_{\mathrm{th}}$, and the
vortex QDs in the current system will be unstable if $N<N_{\mathrm{th}}$.%
{\LARGE \ }In Figs. \ref{character}(b), the blue curves represent vortex QDs
with $S=1$\ is dynamically stable when the norm values exceeding certain
threshold ($N_{\mathrm{th}}\approx 147$). However, in the interval of $N<N_{%
\mathrm{th}}$, a black short dashed curves, which represents the unstable
vortex QDs, is also shown by the figure. Further, Figs. \ref{character}(c,d)
display $N_{\mathrm{th}}$ as a function of $g$ and $S$, respectively. These
thresholds are verified by computing the eigenvalues of $\lambda $, as well
as the direct simulations. It is found that $N_{\mathrm{th}}(g)$ (for $S=1$)
and $N_{\mathrm{th}}(S)$ (for $g=1$) are satisfied, respectively, as
\begin{equation}
N_{\mathrm{th}}^{(S=1)}(g)=\alpha /g,\quad N_{\mathrm{th}}^{(g=1)}(S)=\alpha
S^{2}.  \label{Nth}
\end{equation}%
The numerical simulation demonstrates that $\alpha \approx 147$, which can
be identified by means of the threshold norm at $(S,g)=(1,1)$. According to
our results, the smallest diameter for the vortex QDs, which is $\left(
N,S\right) =\left( 147,1\right) $, is $\sim 20$. If we assume the
confinement is $2\mu m$, according to the above transformations, $x=1\sim
\sqrt{2}\times 2\mu m\approx 2.8\mu m$. Therefore, the real size for this
vortex droplets is $\sim 60\mu m$. If we want to let the 2D solution is
relevant in the 3D system, \textquotedblleft $a_{\perp }$\textquotedblright\
should be at least expand to $>10\mu m$. It is necessary to piont out that
the threshold $N_{th}\left( S\right) $ for the vortex QDs in the current
system is larger than their counterparts in model of thin confinement in
Ref. \cite{Yongyao2018}, which indicates that the stability area of the
vortex QDs for the thin confinement system is larger than the thick
confinement. Therefore, the vortex QDs in the model of thin confinement
should be more stable than the current system.

In the following subsection, we will provide the theoretical analysis with
respect to some of the numerical results in this subsection.

\subsection{Analysis}

Some analyses based on the the characteristics of the vortex QDs are
conducted as follow: because the vortex QDs generally have flat-top density
profiles, one can therefore apply the Thomas-Fermi (TF) approximation to
analyze them, i.e., neglecting the contribution from the kinetic term.
Therefore, the energy density of the soliton can be written as
\begin{equation}
\epsilon (n_{\max })=-{\frac{g}{2}}n_{\max }^{2}+{\frac{2}{5}}n_{\max
}^{5/2},  \label{correct}
\end{equation}%
where $n_{\max }=|\Psi |_{\max }^{2}$ is the peak value of the density
profile. The value of $n_{\max }$ is determined by the minimization of the
total energy. If the radius of the flat-top soliton is $R$, the total norm is%
\begin{equation}
N\approx A_{\mathrm{eff}}n_{\max },  \label{NR}
\end{equation}%
hence the area of the soliton can be written as $A_{\mathrm{eff}}\approx
N/n_{\max }$ (this equation partially explains why the function of effective
area versus $N$, \ i.e., $A_{\mathrm{eff}}\left( N\right) $, can be linearly
fitted) and, accordingly, the total energy is%
\begin{equation}
E\approx A_{\mathrm{eff}}\epsilon (n_{\max })\approx N\left( -{\frac{g}{2}}%
n_{\max }+{\frac{2}{5}}n_{\max }^{3/2}\right) .  \label{E}
\end{equation}
This relationship explains the reason of $E$ linearly depends on $N$ in Fig. %
\ref{Aeff}(f). Finally, the system selects the value of $n_{\max }$ which
minimizes the total energy for fixed $N$: $dE/dn_{\max }=0$, hence
\begin{equation}
n_{\max }=\left( {\frac{5}{6}}g\right) ^{2}.  \label{nmax}
\end{equation}%
If $g=1$, one may obtain $n_{\max }=|\Phi |_{\max }^{2}\approx 0.694$, which
is in accordance with the magnitude of $|\Psi |_{\max }^{2}$ if the limit of
$N$ is sufficiently large in Fig. \ref{character}(a). According to the above
analysis,
\begin{equation}
\mathrm{when}\quad N\rightarrow \infty ,\quad \mu \rightarrow -\left( {\frac{%
5}{6}}g\right) ^{2}g+\left( {\frac{5}{6}}g\right) ^{3}.  \label{mu}
\end{equation}%
For $g=1$, one may obtain $\mu \approx -0.1154$. In Fig. \ref{character}(b),
we can see that the $\mu (N)$ curve also trends to this limit.

The same results for $n_{\max }$ and the respective value $\mu (n_{\max })$
can also be found in a different way. To this end, note that in the limit of
very broad solitons, radial equation (\ref{phir}) becomes
quasi-one-dimensional%
\begin{equation}
\mu \phi =-\frac{1}{2}\frac{d^{2}\phi }{dr^{2}}-g\phi ^{3}+\phi ^{4}.
\label{Q1D}
\end{equation}%
This equation can be derived from a formal Hamiltonian (if $r$ is formally
treated as time), which remains constant in the course of the evolution
along $r$%
\begin{equation}
h=\frac{1}{2}\mu \phi ^{2}+\frac{1}{4}\left( \frac{d\phi }{dr}\right) ^{2}+{%
\frac{g}{4}}\phi ^{4}-{\frac{1}{5}}\phi ^{5}.  \label{h}
\end{equation}%
For solitons, $\phi (r=\infty )=0$, and hence one should set $h=0$ in Eq. (%
\ref{h}).

In the limit of very broad solitons, the derivative terms in Eqs. (\ref{Q1D}%
) and (\ref{h}) may be dropped, which yields an algebraic system%
\begin{equation}
\mu =-g\phi ^{2}+\phi ^{3},~\mu =-{\frac{g}{2}}\phi ^{2}+{\frac{2}{5}}\phi
^{3}.  \label{system}
\end{equation}%
A solution of this system is identical to the values given by Eqs. (\ref%
{nmax}) and (\ref{mu}).

\subsection{Dynamical process}

\begin{figure}[h!]
{\includegraphics[width=0.9\columnwidth]{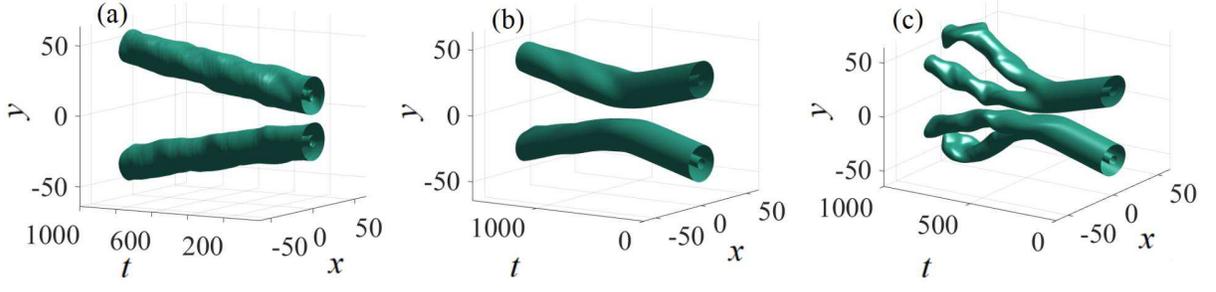}}
\caption{Interactions between two vortex QDs. (a)Automatic repulsion. (b)
Collision with small kick ($\protect\eta=0.025$), which creates a small
relative velocity between them. (c) Collision with strong kick ($\protect\eta%
=0.05$), which gives rise to a large value of relative velocity between
them. Here, the vortex QDs are select with $(N,S,g)=(200,1,1)$.}
\label{dynamic}
\end{figure}

The dynamical process in this subsection is mainly focused on the
interaction between two vortex QDs. Unlike the counterparts in the thin
confinement system, which can merge with each other during the meeting
between two QDs because the logarithm term can feature alternative
attraction and repulsion, two QDs feature strong repulsion between each
other because the LHY term manifests a quartic repulsion under the new
condition. If the two vortex QDs close to each other (i.e., they have some
overlap of their tails), they automatically repel each other. A typical
example for this dynamic is displayed in Fig. \ref{dynamic}(a). In this
figure, two QDs automatically move away from each other because of the
repulsion between them. The repulsive effect also influences the collision
between two moving vortex QDs. If the relative velocity between two
colliding vortex QDs is small, an elastic collision occurs. In this case,
two colliding QDs can retain their vortices after their collision. A typical
example of an elastic collision between two moving vortex QDs is shown in
Fig. \ref{dynamic}(b). If the relative speed between them is large, an
inelastic collision is expected. In this case, two stable vortex QDs are
destroyed during their meeting. After the collision, they will separate into
several fragments. A typical example of an inelastic collision of two vortex
QDs with $S=1$ is shown in Fig. \ref{dynamic}(c).

\begin{figure}[h!]
{\includegraphics[width=0.9\columnwidth]{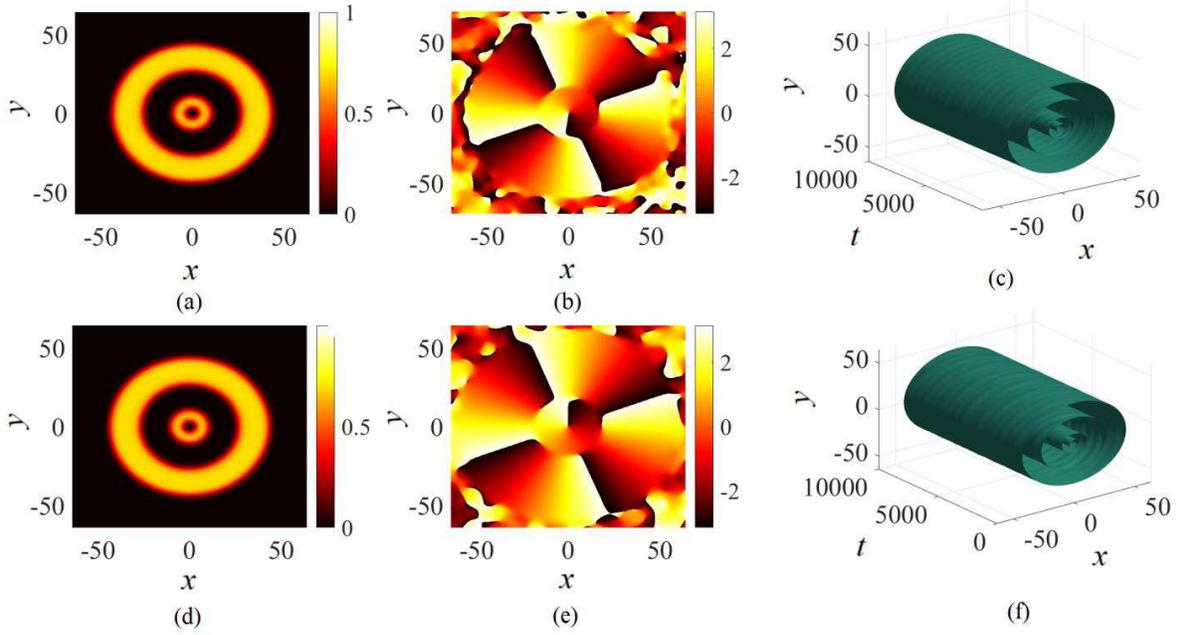}}
\caption{Typical examples of the nested vortex QDs. The upper row includes a
vortex QD with $(N,S,g)=(200,1,1)$ nests inside vortex QD with $%
(N,S,g)=(2500,4,1)$, while the lower row is the vortex QD with $%
(N,S,g)=(200,-1,1)$ nests inside a vortex QD with $(N,S,g)=(2500,4,1)$.
(a,d)The output density pattern for the two nested vortex QDs at $t=10000$.
(b,e)Output pattern of the phase structure for the two nested vortex QDs at $%
t=10000$. (c,f)Direct simulation of the two nested vortex QDs from $%
t=0\rightarrow10000$.}
\label{dynamicnest}
\end{figure}

Because the vortex QD with large topological charge number also has a large
inner radius, a large empty space is created inside the QD. Hence, this
space can be used to embed another QD with a smaller outer radius. If the
outer ring of the embedded QD includes sufficient empty space adjacent to
the inner ring of the encapsulated QD, these two QDs can stable coexist
together for a sufficiently long period of time. Typical examples of such
nested vortex QDs are displayed in Fig. \ref{dynamicnest}. In the examples,
two smaller vortex QDs with $(N,S)=(200,\pm 1)$ are embedded within larger
vortex QDs with $(N,S)=(2500,4)$, respectively. The direct simulations show
that these two nested vortex QDs can stablely coexist beyond $t=10^{4}$.
During the realtime evolution, we can observe small turbulence created on
their density profiles, however, such perturbation cannot destroy the
stability of such a nested configuration.

\section{Conclusion}

The objective of this work is to study the stabilities and characteristics
of 2D vortex quantum droplets (QDs) formed by the Bose-Bose (BB) mixture
under thicker transverse confinement with $a_{\perp }$ of up to at least a
few microns. In this quasi-2D system, the LHY (Lee-Huang-Yang) term is
replaced by its original form in the 3D configuration and features quartic
repulsion. Stable 2D vortex QDs can be found in the current system up to at
least $S=4$. Density profiles and the chemical potentials for the vortex QDs
are systematically studied by numerical simulation and theoretical analysis
throughout the paper. The threshold norms for supporting the stable vortex
QDs are reformulated by the current system. Interactions between the vortex
QDs are also considered. Unlike the QDs formed via the thin confinement
system (i.e., $a_{\perp }\ll 1\mu m$), the strong repulsion induced by the
LHY term makes the vortex QDs repel each other in the current setting.
Elastic and inelastic collisions between two moving vortex QDs are
characterized by exerting different relative speeds. In the case of the
elastic collision, the vortex QDs can maintain their vortices after the
collision. In the case of inelastic collision, the vortex QDs are destroyed
and split into fragments after the collision. Dynamics of the nested vortex
QDs were also considered. This shows that the embedded QD and the
encapsulated QD can stably coexist for sufficiently long duration of time if
they have enough empty space between them. The result in this work may help
to study the stabilization of vortex QDs formed by the BB mixture in the 3D
configuration. Further, this work may help to study the stabilization of
vortex QDs formed in optical lattice \cite{Zhouzheng}. Specially, combining
Abrikosov lattice with the current model maybe a very interesting
generation, which may gain many novel phenomena to the QDs.

\begin{acknowledgments}
We appreciate the useful discussion from Prof. Yongyao Li. This work was
supported by the NNSFC (China) through a grant Nos. 11905032, 11874112, the
Key Research Projects of General Colleges in Guangdong Province through
grant No. 2019KZDXM001, the Foundation for Distinguished Young Talents in
Higher Education of Guangdong through grant No. 2018KQNCX279, and the
Special Funds for the Cultivation of Guangdong College Students Scientific
and Technological Innovation, No. xsjj202005zra01.
\end{acknowledgments}

\end{document}